\documentclass[aps,prl,10pt,twocolumn,amsmath,amssymb,showpacs,superscriptaddress]{revtex4}

\usepackage{subfigure}
\usepackage{epsfig}
\usepackage{float}
\usepackage{multirow}
\usepackage{url}

\usepackage{graphicx}
\usepackage{amssymb, amsmath}
\usepackage[usenames]{color}
\usepackage{dcolumn}
\usepackage{bm}
\usepackage{ textcomp }
\usepackage[normalem]{ulem}
\usepackage{multirow}

\begin{document}

\title{Liganded Xene as a Prototype of Two-Dimensional Stiefel-Whitney Insulators}

\author{Mingxiang Pan}
\author{Dexin Li}
\author{Jiahao Fan}
\affiliation{School of Physics, Peking University, Beijing 100871, China}

\author{Huaqing Huang}
\email[Corresponding author: ]{huaqing.huang@pku.edu.cn}
\affiliation{School of Physics, Peking University, Beijing 100871, China}
\affiliation{Collaborative Innovation Center of Quantum Matter, Beijing 100871, China}
\affiliation{Center for High Energy Physics, Peking University, Beijing 100871, China}

\date{\today}

\begin{abstract}
Two-dimensional (2D) Stiefel-Whitney insulator (SWI), which is characterized by the second Stiefel-Whitney class, is a new class of topological phases with zero Berry curvature. As a novel topological state, it has been well studied in theory but seldom realized in realistic materials. Here we propose that a large class of liganded Xenes, i.e., hydrogenated and halogenated 2D group-IV honeycomb lattices, are 2D SWIs. The nontrivial topology of liganded Xenes is identified by the bulk topological invariant and the existence of protected corner states. Moreover, the large and tunable band gap (up to 3.5 eV) of liganded Xenes will facilitate the experimental characterization of the 2D SWI phase. Our findings not only provide abundant realistic material candidates that are experimentally feasible, but also draw more fundamental research interest towards the topological physics associated with Stiefel-Whitney class in the absence of Berry curvature.
\end{abstract}
\maketitle

\paragraph{Introduction.}---
With the rapid progress of topological states, the concept of Berry curvature and associated topological invariants, such as Chern numbers \cite{PhysRevLett.49.405,PhysRevB.31.3372}, mirror or spin Chern numbers \cite{PhysRevB.78.045426,hsieh2012topological,PhysRevLett.97.036808,PhysRevB.75.121403,PhysRevB.80.125327}, and Fu-Kane invariants \cite{PhysRevB.74.195312,JPSJ.76.053702}, have been widely applied to condensed matter physics. Recently, a new class of topological state with zero Berry curvature, which is characterized by the Stiefel-Whitney (SW) class, was proposed in spinless systems with space-time inversion symmetry $I_{ST}=PT$ or $C_{2z}T$, where $P$, $T$, and $C_{2z}$ are spatial inversion, time-reversal, and two-fold rotation symmetry, respectively \cite{PhysRevLett.121.106403,Ahn_2019, PhysRevX.9.021013}. This is the so-called SW insulator (SWI), which is topologically distinguished by a different topological invariant, i.e., the second SW number $w_2$ \cite{PhysRevB.99.235125}. Different from topological states associated with Chern class which possess topological boundary states due to the bulk-boundary correspondence, a 2D SWI features topologically protected corner states in the presence of additional chiral symmetry, indicating it is also a new class of 2D second-order topological insulators (SOTIs) \cite{benalcazar2017quantized,schindler2018higher, PhysRevLett.120.026801, PhysRevLett.119.246401,PhysRevLett.119.246402}. So far, SOTIs have been proposed in various systems, including crystalline solids and artificial structures \cite{schindler2018bismuth,yue2019symmetry, PhysRevLett.122.256402, PhysRevLett.124.136407,PhysRevLett.123.186401, PhysRevLett.123.256402,lee2020two,liu2019two, PhysRevLett.123.216803,PhysRevLett.126.066401,PhysRevLett.125.056402}. However, the newly proposed 2D SWI was mainly studied in theory but seldom in realistic materials, which greatly hinder the experimental study of SWIs.
It is thus emergent to search for 2D SWIs in realistic materials.

Meanwhile, in the field of 2D materials, a monoelemental class of 2D honeycomb crystals termed Xenes (X refers to C, Si, Ge, Sn and so on) \cite{molle2017buckled,zhao2020two,BECHSTEDT2021100615} have attracted tremendous attention as they provide an ideal platform to explore various topological physics. More than a dozen different topological phases, including the quantum spin Hall (QSH) \cite{PhysRevLett.111.136804,PhysRevB.89.115429,PhysRevLett.107.076802,PhysRevLett.95.226801}, quantum anomalous Hall \cite{PhysRevLett.113.256401}, quantum valley Hall states \cite{PhysRevLett.109.055502,PhysRevB.87.155415,ezawa2015monolayer}, and topological superconductors \cite{PhysRevB.90.054503,PhysRevLett.123.126402, falson2020type, liao2018superconductivity}, are predicted to emerge in Xenes, and these topological states are easily tuned, for example, by chemical functionalization with ligands. In particular, depending on the type of ligands, hydrogenated or halogenated derivatives of Xene can be large-gap QSH or trivial insulators with tunable gaps \cite{PhysRevLett.111.136804,PhysRevB.89.115429}.

In this Letter, we extend the theoretical prediction and experimental applicability of the topological physics associated with SW class by recognizing that the liganded Xene family XL (X=C, Si, Ge, Sn, L=H, F, Cl, Br, I), a large, well-studied, and readily synthesizable class of materials\cite{mannix2017synthesis, Grazianetti2020xenes, ANTONATOS2020100502,zhang2021recent}, are 2D SWIs. Based on first-principles calculation and theoretical analysis, the chemical bonding configuration, bulk topological invariant, and in-gap topological corner states are calculated to identify the SW topology. Moreover, the large and tunable band gaps of liganded Xenes will largely facilitate experimentally observing in-gap corner states. Since some liganded Xenes have been experimentally synthesized, we believe our proposal has strong feasibility to be detected by transport and STM measurements, which may draw immediate experimental attention.

\paragraph{Method.}---
We perform the first-principles calculations within the framework of density functional theory using the Vienna {\it ab initio} simulation package \cite{VASP}. The exchange-correlation functional is treated using the Perdew-Burke-Ernzerhof (PBE) generalized-gradient approximation \cite{PBE}. The predicted topology is further verified by using the Heyd-Scuseria-Ernzerhof (HSE) hybrid functional \cite{heyd2003}. The negligible spin-orbit coupling is ignored in our calculations. We also generate maximally localized Wannier functions (MLWFs) for the analysis of chemical bonding
\cite{wannier90,RevModPhys.84.1419}. 

\begin{figure}
\includegraphics[width=1\columnwidth]{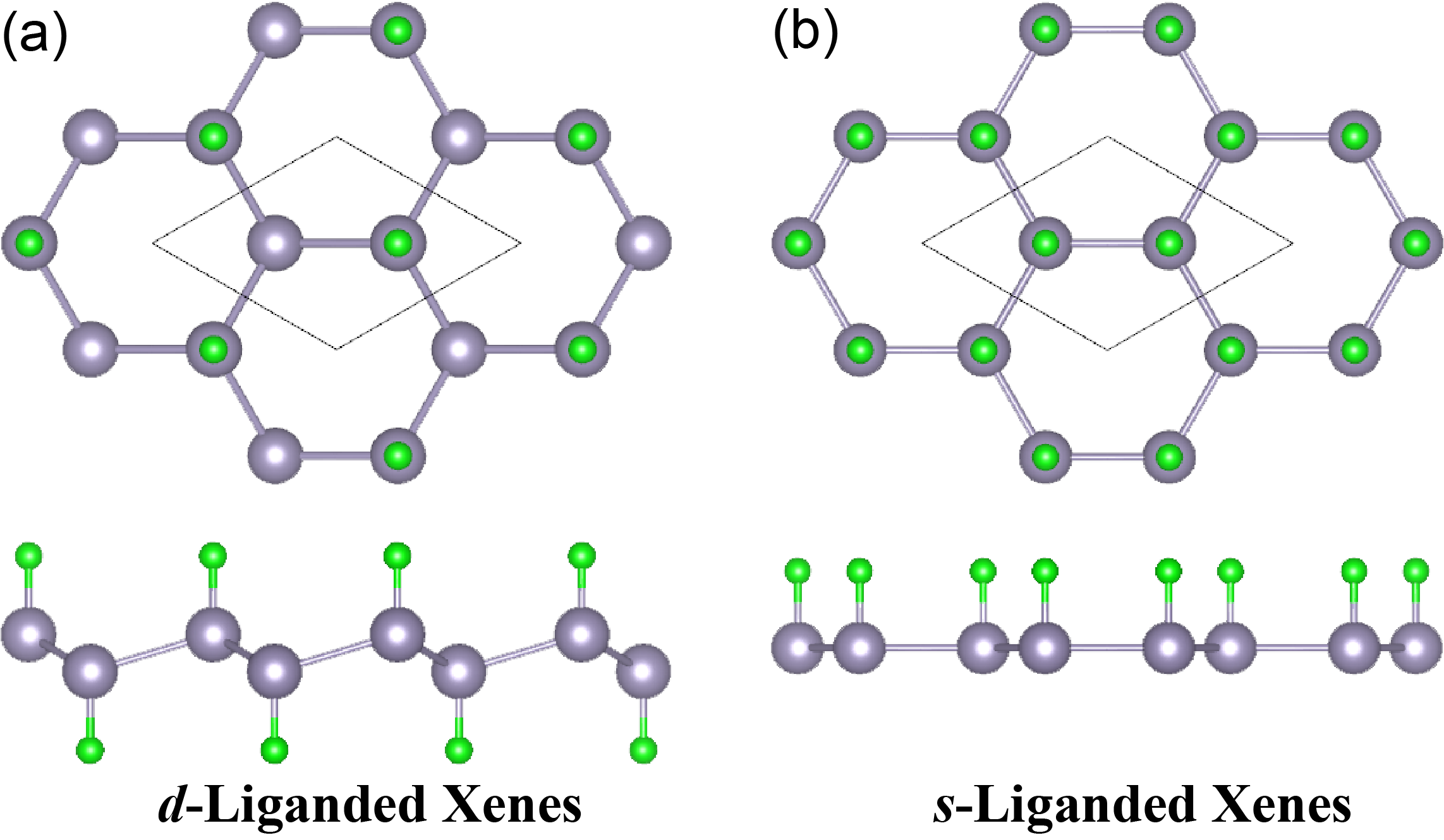} 
	\caption{\label{fig1} (a),(b) Crystal structure for double-side and single-side liganded Xenes (e.g., $d$-CH and $s$-CH) from the top (side) view [upper (lower)].}
\end{figure}

\paragraph{Results.}---
We begin our discussion by introducing the atomic structure and associated crystalline symmetry. Since the compounds in the ligand-terminated Xenes family have similar crystal structures and electronic structures, we take the single-side and double-side hydrogenated graphene (also named \textit{graphane}), denoted as $s$-CH and $d$-CH, as an example hereafter. Figure~\ref{fig1}(a) and~\ref{fig1}(b) show crystal structures for $d$-CH and $s$-CH, respectively. The $d$-CH with H alternating on both sides of the nanosheet is in a buckled hexagonal honeycomb structure with space group 164 ($D^3_{3d}$) including inversion symmetry $P$, while the $s$-CH is in a planar configuration with space group 183 ($C_{6v}^1$)
which contains $C_{2z}$, the two-fold rotation symmetry about the $z$-axis. As time-reversal symmetry $T$ exists in both systems, therefore, the space-time inversion symmetry required for 2D SWIs are $I_{ST}=PT$ and $C_{2z}T$ for $d$-CH and $s$-CH, respectively. The optimal lattice constants are 2.54 and 2.84 \AA~for $d$-CH and $s$-CH respectively, which are consistent with previous reports \cite{PhysRevB.75.153401,PhysRevB.84.041402}. 

\begin{figure}
\includegraphics[width=1\columnwidth]{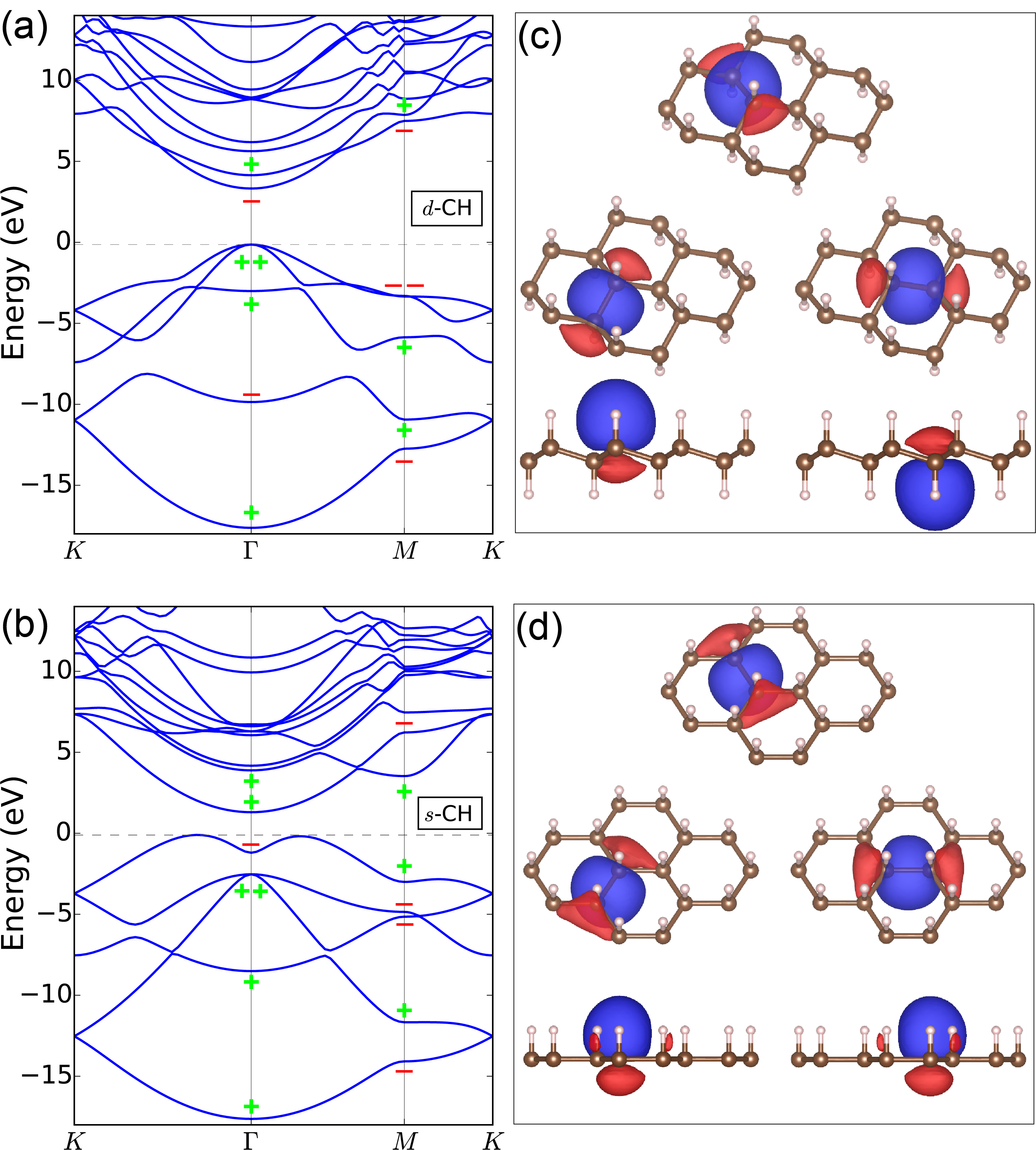} 
	\caption{\label{fig2} Band structure for (a) $d$-CH and (b) $s$-CH without (black solid lines) and with (red dashed lines) spin-orbit coupling. $+/-$ marks Bloch states having opposite parities with respect to inversion or $C_{2z}$ operation at high-symmetry points of the Brillouin zone. Isosurface contours of maximally localized Wannier functions (MLWFs) constructed from the five valence bands of (c) $d$-CH and (d) $s$-CH, displaying the character of $\sigma$-bonded combinations of $sp^3$ hybrids (blue for positive value and red for negative).}
\end{figure}

For the hydrogenated graphene structures, H atoms directly couple to the half-filled $p_z$ orbitals in intrinsic graphene, thereby removing $\pi$ bonding and forming H-C $\sigma$ covalent states. The structural buckling in $d$-CH further enhances a $sp^3$ hybridization of C atoms. There are in total five covalent bonds within the unit cell, including three C-C and two H-C bonds. Since the valence electron configuration of C and H are $2s^22p^2$ and $1s^1$, covalent bonding states are fully occupied with two electrons per bonding state. Therefore, the system tends to be an insulator with the Fermi level lying in the gap between bonding and antibonding states. As shown in Fig.~\ref{fig2}(a) and \ref{fig2}(b), $d$-CH ($s$-CH) is indeed an insulator with a direct (indirect) gap. Interestingly, the band gap of $s$-CH ($\sim1.4$ eV) is much smaller than that of $d$-CH ($\sim3.46$ eV). Although the lack of structural corrugation in $s$-CH leads to deviations away from the $sp^3$ hybridization, single-side hydrogenation brings H atoms closer together, which naturally results in a much larger repulsion among the H-C $\sigma$ bonding states. As a consequence, the occupied band derived from H-C bonds [marked by ``-'' at $\Gamma$ in Fig.~\ref{fig2}(b)] shifts upwards in energy that gives rise to a smaller indirect band gap of $s$-CH.

To get a better understanding of the hybridization, covalency, and ionicity of chemical bonds in $d$- and $s$-CH, we construct MLWFs from the five occupied valence bands. As shown in Fig.~\ref{fig2}(c) and ~\ref{fig2}(d), it is clear that these MLWFs display the character of $\sigma$-bonded combinations of mixed $sp^2$-$sp^3$ hybrids, providing an intuitive chemical signature of covalent bonds. Moreover, it can be seen that the MLWFs of H-C covalent bonds and their Wannier charge centers are shifted towards H due to the difference of electronegativity between H and C.

The common covalent $\sigma$ bonding states have the average charge center located around the middle of the bond, implying the unconventional feature of the mismatch between average electronic centers and atomic positions (also known as obstructed atomic limits \cite{PhysRevB.97.035139,PhysRevB.99.245151}). This is a direct indicator of nontrivial band topology. We then perform a band representation (BR) analysis based on the theory of topological quantum chemistry \cite{bradlyn2017topological,vergniory2019complete,tang2019comprehensive, zhang2019catalogue}. The BR of trivial atomic insulators is solved to be a sum of atomic-orbital-induced BRs (aBRs), while that of unconventional materials, such as SOTIs, must be a combination of some aBRs and an essential BR from an empty Wyckoff position where no atom exists \cite{gao2021unconventional,PhysRevB.103.205133}. The BR decomposition of $d$-CH is $A_1@2d+A_g@3e$. The C atoms are located at $2d$ site of space group 164, while $3e$ sites of the essential BR are the C-C bond centers, which correspond to $h_{3c}^{(\bar{3})}$ primitive generator class of SOTIs in the Benalcazar {\it et al.}'s notation \cite{PhysRevB.99.245151}. This implies that there is no net dipole in the plane and the corner charge fractionalization will be $e/2$ in each $\pi/6$ sector. Similarly, we found the BR decomposition of $s$-CH is $A_1@2b+A_1@3c$, also indicating its nontrivial bulk topology (see Supplemental Materials \footnote{\label{fn} See Supplemental Material at http://link.aps.org/supplemental/xxx, for more details about the computational method, Wilson loop, theoretical analysis, and numerical results, which includes Refs.~\cite{VASP,PBE,wannier90,RevModPhys.84.1419,yuruiZ2,PhysRevX.9.021013, bradlyn2017topological,vergniory2019complete,tang2019comprehensive, zhang2019catalogue, gao2021unconventional, PhysRevB.97.035139,PhysRevB.99.245151}.}).

Physically, the nontrivial bulk topology of $d$- and $s$-CH can also be understood from the intuitive picture of double band inversion \cite{PhysRevLett.121.106403,PhysRevLett.123.186401,Hsu13255,xiao2021first,huang2021generic}. For $d$-CH with inversion symmetry $P$, we consider $N_{\mathrm{occ}}^-(\Pi)$, the number of occupied bands with negative parity at time-reversal invariant momenta (TRIM, $\Pi=\Gamma$ and three $M$ points). It is found that $N_{\mathrm{occ}}^-(M)-N_{\mathrm{occ}}^-(\Gamma)= 2$, as shown in Fig.~\ref{fig2}(a). This indicates a double band inversion, because the system cannot be adiabatically connected to the trivial atomic insulator limit where the parity representations at TRIM must be the same. Importantly, the second-order band topology of 2D inversion-symmetric spinless systems can be characterized by the second SW number $w_2$ \cite{PhysRevB.99.235125,Ahn_2019,lee2020two,PhysRevLett.121.106403,PhysRevLett.123.216803}, which is determined by
\begin{equation}\label{stiefel-whitney}
  (-1)^{w_2}=\prod_{\Pi\in \mathrm{TRIM}} (-1)^{\lfloor N_{\mathrm{occ}}^-(\Pi)/2\rfloor},
\end{equation}
where $\lfloor\cdot\rfloor$ is the floor function.
Therefore, the double band inversion gives rise to a nontrivial $w_2=1$, demonstrating that it belongs to the nontrivial SW class. Alternatively, $w_2$ for $s$-CH with $C_{2z}T$ symmetry can be obtained by tracing the Wilson loop spectra (see Supplemental Materials \footnotemark[\value{footnote}]). It is proved that $w_2$ is given by the parity of the number of spectral crossing at $\Theta=\pi$, where $\Theta$ indicates the phase eigenvalue of the Wilson loop operator \cite{PhysRevLett.121.106403,Ahn_2019,PhysRevX.9.021013}. Due to a similar double band inversion with respect to $C_{2z}$ in $s$-CH, as shown in Fig.~\ref{fig2}(b), we found a nontrivial $w_2=1$, confirming it is a SWI.

To explicitly identify the second-order topology in $d$- and $s$-CH, we calculate the fractional corner charge $Q_{\mathrm{corner}}$, which is a bulk topological index for classifying SOTIs \cite{PhysRevB.99.245151}. In 2D insulators with $C_{6}$ and $T$ symmetries (e.g., $s$-CH), it can be evaluated as \cite{PhysRevB.99.245151}
\begin{equation}
Q_{\mathrm{corner}}=\frac{e}{4}[M_1^{(2)}]+\frac{e}{6}[K_1^{(3)}] \; \mod e,
\end{equation}
where $[\Pi_p^{(n)}] = \#\Pi_p^{(n)} -\#\Gamma_p^{(n)}$ and $\#\Pi_p^{(n)}$ is defined as the number of bands below the band gap with $C_n$ rotation eigenvalues $\Pi_p=\exp[\frac{2\pi i (p-1)}{n}]$ for $p = 1, 2, \cdots, n$. $\Pi$ stand for high symmetric point $M$ and $K$. For $d$-CH with $S_6$ symmetry, the above formula should be modified by replacing $[M_1^{(2)}]$ with $[M_\pm^{(i)}]$, which is the difference in the number of bands with inversion eigenvalue even/odd between $M$ and $\Gamma$ \cite{PhysRevB.102.115104}. Based on the first-principles calculations, we have $[M_1^{(2)}]=-2$, $[K_1^{(3)}]=0$ for $s$-CH and $[M_\pm^{(i)}]=\pm 2$, $[K_1^{(3)}]=0$ for $d$-CH. Therefore, $Q_{\mathrm{corner}}={e}/{2}$, indicating both $s$-CH and $d$-CH to be 2D SOTIs \cite{PhysRevResearch.1.033074}.

\begin{figure*}
\includegraphics[width=0.85\textwidth]{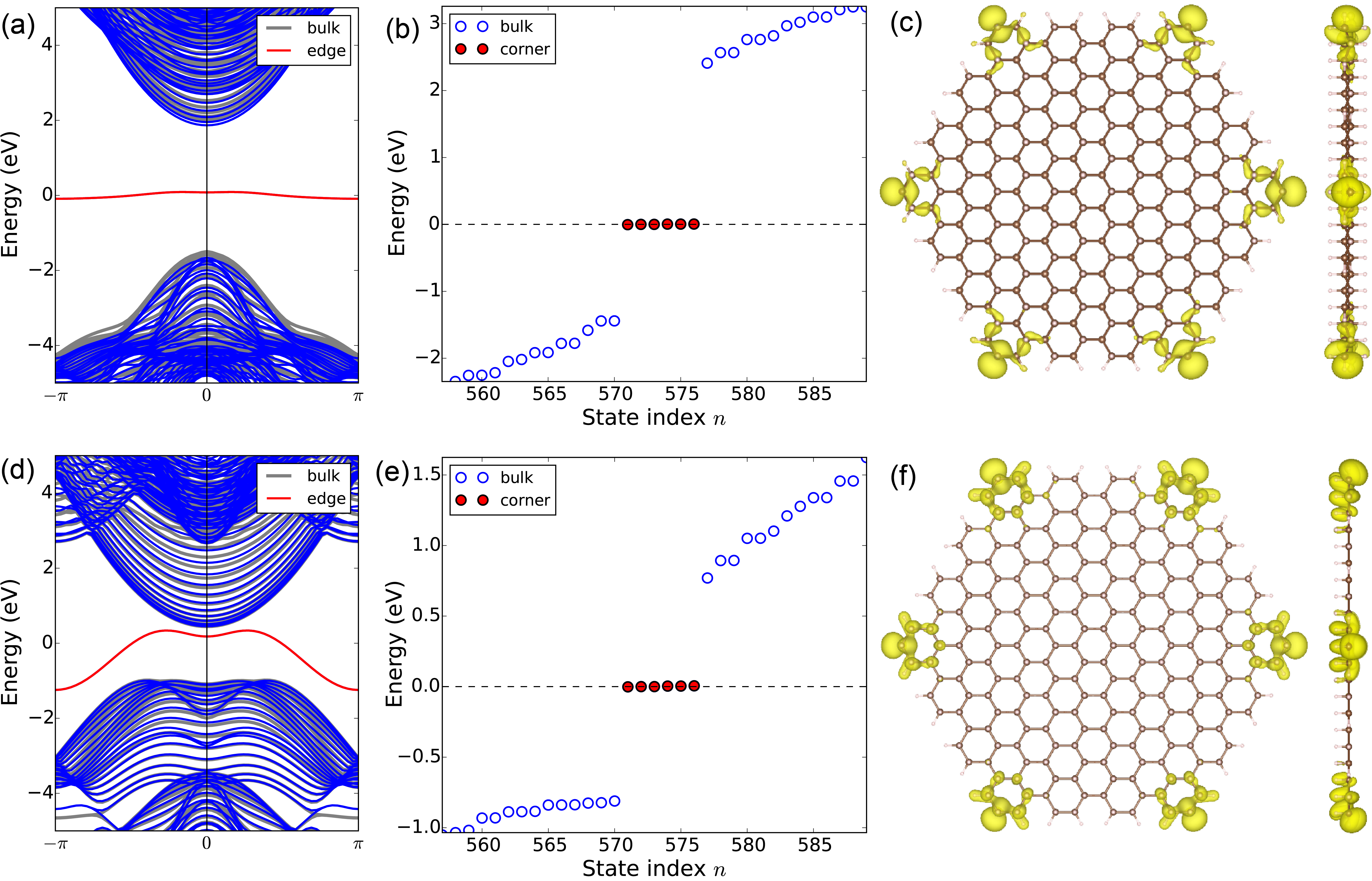} 
	\caption{\label{fig3} SWI in (a-c) $d$-CH and (d-f) $s$-CH honeycomb monolayer. (a,d) Band structure of a nanoribbon of hydrogenated graphene with (gray) and without (blue and red) hydrogen saturation. The flat edge bands are marked in red. (b,d) The energy spectrum of a hexagonal-shaped nanodisk with H-saturated edges. (c,f) Top and side view of the real-space charge distribution of corner states around the Fermi level.}
\end{figure*}

To further reveal their topological nature, the first-principles calculations are performed to directly verify the gapped topological edge states and in-gap topological corner states in $d$- and $s$-CH, which can be used as a fingerprint to distinguish them from other topological phases. As shown in Fig.~\ref{fig3}(b) and~\ref{fig3}(d), an edge band appears throughout the gap and detaches from bulk states of a zigzag nanoribbon. Since the first SW number $w_1$, which is equivalent to the Zak phase, is turned out to be trivial ($w_1=0$), the existence of the 1D edge state is not due to the bulk polarization, but it is more like a dangling bond state (see Supplemental Materials \footnotemark[\value{footnote}]). 
In fact, different from the topologically protected helical edge states of 2D topological insulators, these edge states are less robust and can be removed by saturating the dangling bonds at edges with hydrogen atoms. Similar flat edge states also occur in the monolayer $\beta$-Sb, which has been proved to be a 2D SOTI \cite{PhysRevB.102.115104,PhysRevResearch.1.033074}.

Due to the modified bulk-boundary correspondence, the most direct indicator of SOTIs is the corner-localized in-gap states. To identify the corner topology of $d$- and $s$-CH, we calculate the energy spectrum for hexagonal nanodisks with $\sim 500$ atoms. In order to remove dangling-bond edge states from the bulk gap, we passivate the edges with hydrogen atoms \cite{PhysRevResearch.1.033074}. The energy spectrum for the nanodisk of $d$-CH and $s$-CH are plotted in Fig.~\ref{fig3}(b) and ~\ref{fig3}(e), respectively. Remarkably, there are six states that are degenerate at the Fermi level inside the bulk gap. Moreover, an emergent chiral (sublattice) symmetry, which derives from the bipartite honeycomb lattice, pins the corner modes in the mid of the gap. 
As shown in Fig.~\ref{fig3}(e) and ~\ref{fig3}(f), these states are well localized at six corners of the hexagonal sample, confirming that they are the topological corner states. At exact half-filling, three of the six states are occupied, leading to a fraction corner charge of $Q_{\mathrm{corner}}=e/2$ per corner which is consistent with the above analysis.

\begin{table}
\centering
	\caption{\label{tab1} Topological phases of single-side ($s$-) or double-side ($d$-) liganded Xene. The band gaps ($E_g$) are obtained from first-principles calculations or refer to experiments. $w_2/Z_2$ denotes the type of topological states (i.e., the SWI or QSH phase). 
}
\begin{tabular}{l|ccc|ccc}
  \hline
  \hline
   &  & $s$- &  &   & $d$- & \\ 
  \hline
   & $w_2/Z_2$ & $a$ (\AA) & $E_g$ (eV) & $w_2/Z_2$  &$a$ (\AA) &$E_g$ (eV)\\ 
  \hline
  CH  & SWI   & 2.84&1.40 \cite{PhysRevB.84.041402} & SWI &  2.54&3.47 \cite{balog2010bandgap,PhysRevB.75.153401} \\
  CF  & SWI   & 3.51&0.70  & SWI & 2.60&  3.09 \cite{nair2010fluorographene}\\
  \hline
  SiH  & SWI   & 4.15&1.86 & SWI & 3.89&2.19 \\
  SiF  & SWI  & 4.32&1.30 & SWI & 3.95&0.68 \\
  SiCl & QSH  & 5.00&0.02 & SWI & 3.94&1.28 \\
  SiBr & QSH  & 5.45&0.05 & SWI & 3.97&1.20 \\
  SiI  & QSH  & 6.15&0.09 & SWI & 4.06&0.55 \\
  \hline
  GeH  & SWI   & 4.36&1.57 & SWI & 4.08&0.98 \cite{bianco2013stability}\\
  GeF  & QSH  & 4.64&0.11 & SWI & 4.22&0.17 \cite{BECHSTEDT2021100615,PhysRevB.89.115429}\\
  GeCl & QSH  & 5.29&0.12 & SWI & 4.24&0.37 \cite{BECHSTEDT2021100615,PhysRevB.89.115429}\\
  GeBr & QSH  & 5.67&0.14 & SWI & 4.25&0.06 \cite{BECHSTEDT2021100615,PhysRevB.89.115429}\\
  GeI  & QSH  & 6.28&0.17 & QSH& 4.32&0.30 \cite{BECHSTEDT2021100615,PhysRevB.89.115429} \\
  \hline
  SnH  & SWI   & 4.99&1.32 & SWI & 4.71&0.47 \\
  SnF  & QSH  & 5.24&0.24 & QSH &5.01 &0.29 \cite{PhysRevLett.111.136804} \\
  SnCl & QSH  & 5.58&0.24 & QSH &4.93 &0.26 \cite{PhysRevLett.111.136804} \\
  SnBr & QSH  & 5.84&0.25 & QSH &4.91 &0.29 \cite{PhysRevLett.111.136804} \\
  SnI  & metal & 6.06& 0  & QSH &4.90 &0.34 \cite{PhysRevLett.111.136804} \\
  \hline
  \hline
\end{tabular}
\end{table}

In addition, we also investigated other liganded Xenes. They are counterparts of hydrogenated graphene, corresponding to the silicene, germanene, and stanene monolayer saturated by hydrogen or halogen. Some of these materials, such as hydrogenated graphene  \cite{elias2009control,balog2010bandgap,REVgraphane} and germanene \cite{bianco2013stability}, or fluorinated graphene (also named fluorographene), have been experimentally synthesized \cite{jeon2011fluorographene,nair2010fluorographene,zbovril2010graphene, bianco2013stability}. Their topological properties have been carefully investigated according to the conventional classification of time-reversal $Z_2$ topology. Owing to their negligible spin-orbit coupling, some liganded Xenes are identified as topologically trivial. However, this argument does not forbid an SW topology with zero Berry curvature.

Based on systematic calculations, we found that those prior trivial liganded Xenes are SWIs actually (see Table 1). Since the $\pi$ orbitals are saturated by H or F, these compounds become insulators with band gaps in a wide range. Remarkably, the band gaps of single-side hydrogenated germanane ($s$-GeH) and stanene ($s$-SnH) are larger than that of double-side hydrogenated ones, which is different from that of hydrogenated graphene. This is because the H-X $\sigma$ level does not shift to the valence band maximum due to the weaker repulsion between these states at large distances. Furthermore, among halogenated Xenes, more electronegative ligands (e.g., from F to I in halogens) tend to withdraw electron density from the Xene framework, and lower the energy of the ligand-X $s$-orbital antibonding levels at the conduction band minimum \cite{PhysRevLett.70.1116,molle2017buckled}. Therefore, the trend of band-gap reduction and the topological phase transition from SOTI to QSH are observed from F to I in both single and double-side halogenated Xenes.

Finally, for experimental detection on 2D SWIs, it is preferred to have the corner states sitting deep in the bulk gap. Firstly, the band gaps of liganded Xenes are large and tunable by the saturation position ($s$- or $d$-), the ligand type (hydrogen or halogens), and external perturbations such as electric field and strain. Secondly, the emergent approximate chiral (sublattice) symmetry of the honeycomb lattice structure pins the corner states approximately in the middle of the gap. All these features will facilitate the experimental characterization of the SWI phase in liganded Xenes. Moreover, given that some candidate materials have already been synthesized successfully in experiments \cite{mannix2017synthesis,Grazianetti2020xenes,ANTONATOS2020100502,zhang2021recent,PhysRevX.7.041069}, the SWI phase is highly accessible, and may already be realized in existing materials.

\paragraph{Conclusion.}---
In conclusion, we have demonstrated the 2D SWI phase in a large class of hydrogenated and halogenated Xenes that are experimentally feasible. The nontrivial topological nature of these materials is identified through the nontrivial second SW number $w_2=1$ and the existence of in-gap topological corner states. The candidate materials with band gaps as large as 3.5 eV will facilitate experimentally detecting in-gap corner states by STM measurements. Our results enrich the topological physics associated with SW class, and greatly extend the territory of candidate materials for 2D SWIs. In addition, it is also possible to realize 3D weak and strong SWIs by stacking these 2D SWI candidate materials. These new discoveries may draw more fundamental research interests of Xenes, and provide a practical avenue for the realization of SWIs in real materials that are experimentally feasible.

\begin{acknowledgments}
We thank Jiaheng Gao and Zhijun Wang for the help in determining the BR decomposition. This work was supported by the National Natural Science Foundation of China (Grant No. 12074006) and the start-up fund from Peking University. The computational resources were supported by the high-performance computing platform of Peking University.
\end{acknowledgments}

\paragraph{Note added.} After submission, we become aware of an independent work on arXiv recently\cite{qian2021second}, where the results of liganded Xenes are consistent with ours.

\providecommand{\noopsort}[1]{}\providecommand{\singleletter}[1]{#1}%

\end{document}